# *On Triangle Inequality Based Approximation Error Estimation*

A.K. Alekseev[1*], A.E. Bondarev[2], I. M. Navon[3]

[1]*Moscow Institute of Physics and Technology, Moscow, Russia*
[2]*Keldysh Institute of Applied Mathematics RAS, Moscow, Russia*
[3]*Department of Scientific Computing, Florida State University, Tallahassee, FL 32306-4120, USA*

## *Abstract*

The ensemble of solutions, generated by different solvers, is analyzed from the viewpoint of the approximation error estimation. The distance between the true and numerical solutions in some metric is considered as the approximation error magnitude. The analysis of distances between the numerical solutions provides an opportunity for the estimation of the error magnitude upper bound. Numerical tests for the steady supersonic flows, governed by the two-dimensional Euler equations, are conducted to demonstrate the estimation of the approximation error magnitude in different metrics.

## *1. Introduction*

The modern abundance of different numerical methods provides some additional opportunities for the analysis. From this viewpoint, we consider the distances between approximate solutions in the set of metrics. These distances are caused by the approximation (discretization) errors, so, some structure may be induced in the ensemble of numerical solutions by differences in methods, for example, by the order of approximation. The order of approximation of a finite-difference/finite volume scheme is related to the truncation error order. For the system of PDE, formally denoted herein as $Au = f$, the truncation error $\delta u$ may be obtained via Taylor series decomposition of the discrete operator $A_h u_h = f_h$.

The truncation error asymptotic dependence on the spatial step $h$ is usually written as $\delta u = O(h^n)$, where the order $n$ is equal to the minor order of series terms.

The truncation error causes the approximation error $\Delta u = u_h - u$ that can be described by the formal solution $\Delta u = A_h^{-1} \delta u$. For linear problems, the approximation error asymptotics $\Delta u = O(h^n)$ has the same order $n$ (Lax theorem, [1]) if the discrete operator is well-posed (the inverse operator is uniformly bounded $\left\| A_h^{-1} \right\| < C$, the grid is regular and uniform). For the case of nonlinear equations [2-7] the approximation error order is essentially local and varies significantly depending on the type of flow structures. In this case, the observed order of convergence is not equal to the nominal order even in the asymptotic range. A similar situation may be caused by discontinuities in the coefficients [8] or by the irregularity of the grid [9]. Also, there is no convergence, if the discrete operator is not well-posed, for example, this may occur for the Kelvin-Helmholtz instability [10].

*A- priori* estimation of the approximation error norm may be expressed in the form $\left\| \Delta u \right\| < C \cdot h^n$, which contains unknown constant, independent of the numerical solution. It is a standard approach to error analysis in the design of numerical algorithms. *A priori* error estimation also justifies the common practice to stop mesh refining when the dependence of numerical solution on the step size becomes unobservable.

*corresponding author
Email addresses: [1]alekseev.ak@phystech.edu, [2]a_bond2001@mail.ru, [3]navonim@gmail.com

*A- posteriori* estimation of the approximation error [11,12,13] may be presented in the form $\|\Delta u\| \leq \eta_h(u_h)$, where the computable error indicator $\eta_h(u_h)$ depends only on the approximate solution $u_h$. Sometimes [14], the form $\|\Delta u\| \leq C(u_h) \cdot \eta_h$ is used, where $C(u_h)$ is a computable constant, which depends on the numerical solution, and $\eta_h$ is the computable bound of the truncation error (residual) $\|\delta u\| \leq \eta_h$. At present, the best results in *a- posteriori* error estimation are achieved for elliptic equations and finite element methods starting from the work by Babushka [11]. In the finite element notations $\delta u$ corresponds to residual $r_h$ and the approximation error $\Delta u = u_h - u$ is usually noted as $e_h$. In most of practical applications the constant $C(u_h)$ is not estimated, while the error indicator is used for the mesh adaptation. However, *a posteriori* error estimation may provide more general information regarding both the error and the location of the exact solution. For example, [14] demonstrated that the estimation of the stability constant and the residual may be used for the determination of a vicinity of the numerical solution, which contains the exact solution.

Global (approximation, discretization) error $\Delta u$ calculation methods are surveyed in [15,16]. Herein, we discuss some of these approaches, which are most widespread.

The Richardson extrapolation (RE) [17-21] is one of the most popular methods for error estimation and the most natural approach from the standpoint of grid convergence. It enables to determine the refined solution and the error estimate using the set of solutions computed for different meshes. The solutions should belong to the asymptotic range (the error terms of the single order should dominate). Unfortunately, in CFD problems containing discontinuities, the error order on different flow structures varies [2-7]. It causes the application of the mixed order RE (accounting for errors of two different orders) or the generalized RE (accounting for the local order of convergence) [20,21]. However, both the mixed order and the generalized RE (GRE) are not so easy from the computational standpoint as the usual RE. For example, GRE exhibits instabilities at small (close to zero) orders of approximation [20,21], specific for undisturbed domains in flow. Mixed order RE needs for exact values of two leading orders, which are unknown in general. Additionally, the mixed order RE and GRE require three mesh levels in the asymptotic range that implies computations at one additional mesh level at least. Thus, minimum four consequent mesh levels are necessary for RE application to flows with discontinuities that causes extremely high requirements to computer resources.

There are approaches to the estimation of the discretization error based on some presentation of $\delta u$. The truncation error may be computed by the action of a high order scheme stencil on a precomputed flow field [22, 23], by the action of the differential operator on the interpolation of the numerical solution [24], or via a differential approximation [25, 26]. In the simplest option, the estimation of approximation error may be performed using the defect correction [22, 27, and 28]. In the defect correction frame, the truncation error $\delta u$ is used as the source term intended for the correction of solution. The estimation of the approximation error may be performed also via the linearized problem [28], complex differentiation [29] or by adjoint equations [23, 24, 26, 30]. Usually, adjoint equations are applied to the estimation of a valuable functional (drag, lift etc.) uncertainty. Nevertheless, the variant of adjoint method, described in [26], enables estimation of the norm of solution error. Unfortunately, it implies the solution of many adjoint problems that is proportional to the number of grid nodes that implies an extremely high computational burden.

The numerical enclosure [31,32] also provides the feasibility for single-grid rigorous estimations of error (similar to [14]), however, it is essentially based on the residual estimation and the upper bound for the norm of inverse linearized operator. Herein, this approach is not employed.

The general feature of the residual-based methods is the incompleteness of truncation error estimates. For example, the differential approximation methods based on Taylor series [26] do not account for high order terms of the expansion, the postprocessor based methods do not account for the higher scheme truncation errors [23] or the interpolation errors [24].

In the present paper we consider a single-grid analysis of non-intrusive type. The ensemble of calculations, performed by the solvers of different approximation order, is used in order to search the approximation error. Herein, the truncation error is accounted completely, although implicitly, since the analysis is conducted in the space of numerical solutions, which are functions of the total truncation error.

In contrast to the more widespread norm oriented techniques, the current analysis is based on the ensemble of distances (distance matrix) in different metrics [33], that provides a more general and flexible analysis. The norm oriented variant of this approach is presented in [34, 35]. The paper [35] also contains an information on application of metrics and on aspects of exact solution existence nearby the approximate solution.

The Multidimensional Scaling (MDS) [36] concerns formally similar problems, however, we consider the cases when MDS cannot be applied, since the data vector (numerical solution) length greatly exceeds the number of data vectors.

Numerical tests demonstrated that various metrics have significantly different properties from the error estimation perspective. The best characteristics are observed for the IMED metric [37].

The paper is organized as follows. In Section 2 we discuss the opportunities for the discretization error estimation that are provided by *a priori* information regarding error magnitude rating. Section 3 considers *a posteriori* analysis of error relations provided by the ensemble of numerical solutions performed by different solvers. The supersonic flows with dicontinuities, described by the two-dimensional Euler equations, are considered as test problems in Section 4. Section 5 presents the set of metrics, which are used for comparing of solutions. In Section 6 we present the results of the ensemble based error magnitude estimation, performed using different metrics, in comparison with the true error. The solvers, used for the computations are listed in this section. Several issues, concerning the applications of the metric based error analysis, are surveyed in Section 7. Conclusions are presented in the final Section 8.

## *2. A posteriori error estimation for approximate solutions with ranged errors.*

Let's consider the ensemble of numerical solutions obtained on the same uniform grid using finite volume schemes of different approximation order. We denote the numerical solution as the vector $u^{(i)} \in R^N$ ($i$ is the scheme number, $N$ is the number of grid points). The values of an unknown exact solution at centers of cell ("exact" solution) are denoted as $\widetilde{u} \in R^N$. The approximation error magnitude is treated as the distance between the exact and approximate solutions $d(u^{(k)}, \widetilde{u}) = \delta_{0,k}$ in some metric (for example, $d(u^{(k)}, \widetilde{u}) = \left\| u^{(k)} - \widetilde{u} \right\|_{L_2}$).

Let the relation of these approximation error values be known *a priori*. The following theorem may be stated for two numerical solutions $u^{(1)}$ and $u^{(2)}$ having the errors relation $\delta_{0,1} \geq 2 \cdot \delta_{0,2}$.

**Theorem 1.** *Let the distance $\delta_{1,2} = d(u^{(1)}, u^{(2)})$ between two numerical solutions $u^{(1)} \in R^N$ and $u^{(2)} \in R^N$ be known and distances (unknown) between the numerical and exact solutions be related as*

$$\delta_{0,1} \geq 2 \cdot \delta_{0,2}, \tag{1}$$

*then the error of more accurate solution $u^{(2)}$ has the upper bound:*

$$\delta(u^{(2)}, \widetilde{u}) \leq \delta_{1,2} \tag{2}$$

***Proof.*** The triangle inequality [38] for distances $\delta_{0,1}, \delta_{1,2}, \delta_{0,2}$ between points $u^{(1)}, u^{(2)}, \tilde{u}$ may be presented as $\delta_{0,1} \leq \delta_{1,2} + \delta_{0,2}$ or $\delta_{0,1} - \delta_{0,2} \leq \delta_{12}$. By accounting (1) as $\delta_{0,1} - \delta_{0,2} \geq \delta_{0,2}$, one obtains $\delta_{0,2} \leq \delta_{0,1} - \delta_{0,2} \leq \delta_{12}$ and, finally, the desired expression $\delta_{0,2} \leq \delta_{12}$.

The *Theorem 1* may be stated in a slightly more general form: if two solutions are ranged by the error magnitude as

$$\delta_{0,1} > (1+\alpha)\delta_{0,2}, \alpha > 0, \qquad (3)$$

then $\delta_{0,1} - \delta_{0,2} > \alpha \delta_{0,2}$, $\alpha \delta_{0,2} < |\delta_{0,1} - \delta_{0,2}| \leq \delta_{1,2}$ and

$$\delta_{0,2} < \delta_{1,2}/\alpha. \qquad (4)$$

This means that two numerical solutions, having the error relation $1+\alpha, (\alpha > 0)$ in some metric, define the error majorant $\delta_{0,2} < \delta_{1,2}/\alpha$.

So, the distance between two numerical solutions enables the estimation of the error upper bound, if the relation of errors' magnitudes is known *a priori* in some metrics.

### *3. A posteriori analysis of the error magnitude relations*

Despite the widespread opinion that schemes of higher order are more accurate, the evident weakness of *Theorem 1* is the assumption of the existence of solutions with *a priori* ranged error magnitude. For this reason, we consider some options for *a posteriori* check of error rating.

Naturally, the precise solutions are located in a smaller boll around the exact solution, if compare with imprecise ones. *Theorem 1* is justified for the modest condition (1) $\delta_{0,1} > 2\delta_{0,i}^{\max}$. The separation of the distances between approximate solutions into clusters may be considered as providing evidence of error ranging. So, the quantitative criterion, based on dimension of clusters and the distance between them, is of interest.

Let us compare the set of distances $\delta_{1,j}$ and $\delta_{i,j}$, where $u^{(1)}$ is maximally incorrect solution and $u^{(i)}$ is some more accurate solution, $\delta_{0,i}^{\max}$ is the maximum error in the subset of accurate solutions. The maximum of $\delta_{i,j} (i \neq 1)$ (the distance from zero to maximum error in the cluster of accurate solutions) is noted as the upper bound of the accurate solutions' cluster $d_1$, the minimum of $\delta_{1,j}$ is noted as the low bound of the second cluster $d_2$.

The following heuristic criterion may be used in applications instead the *Theorem 1*:

***Conjecture 1***: *If the set of distances between solutions is split into clusters and the distance between clusters is greater than the size of the cluster of more accurate solutions: $d_2 - d_1 > d_1$, then the error of $u^{(i)}$ is majorized by $\delta_{i,1}$: $\delta_{0,i} \leq \delta_{i,1}$, where $u^{(i)}$ belongs to the cluster of more accurate solutions and $u^{(1)}$ is the maximally inaccurate solution.*

This conjecture is based on the inexact assumptions that the dimension of the accurate cluster is equal to $d_1 = 2\delta_{0,i}^{\max}$ $(i \neq 1)$, and the cluster of inaccurate solutions belongs to the interval $(\delta_{0,1} - \delta_{i,\max}, \delta_{0,1} + \delta_{i,\max})$, so $d_2 = \delta_{0,1} - \delta_{0,i}^{\max}$. Since both these evaluations correspond to collinear vectors of error, they are overestimated. If one assumes them to be valid, the relation of accurate cluster dimension and the distance between clusters has the form $\delta_{0,1} - \delta_{0,i}^{\max} > 4\delta_{0,i}^{\max}$. This leads to the relation $\delta_{0,1} > 5\delta_{0,i}^{\max}$, which ensures the condition (1) $\delta_{0,1} > 2\delta_{0,i}^{\max}$. Formally, in this frame, the relation $d_2 - d_1 > d_1/4$ provides exactly $\delta_{0,1} > 2\delta_{0,i}^{\max}$, however, numerical tests provide the success only for $d_2 - d_1 > d_1$.

It should be noted, that *Conjecture 1* implies no assumptions on the asymptotic range or any convergence of solutions. Formally one may write $\delta_{0,1}/\delta_{0,i} = C^{(1)}/C^{(i)} h^{n_1 - n_i}$ for the asymptotic range and $n_i > n_1$. However, one cannot rely on this asymptotic, since the relation $n_i \approx n_1 \sim 1$ [2-7] is standard for the problems with discontinuities.

This criterion may be rigorous only in the limit of the infinite set of solutions obtained by independent methods. Nevertheless, numerical tests for two dimensional supersonic inviscid flows demonstrate that the collection of distances between solutions $\delta_{i,j}$ enables a detection of the close and distant solutions, if the error magnitudes are significantly different. The numerical tests confirm the applicability of this heuristic criterion, however, with a significant dependence on applied metrics.

### *4. Test problems*

The tests problems are related with several flow patterns, governed by two dimensional Euler equations

$$\frac{\partial \rho}{\partial t} + \frac{\partial (\rho U^k)}{\partial x^k} = 0 ; \tag{5}$$

$$\frac{\partial (\rho U^i)}{\partial t} + \frac{\partial (\rho U^k U^i + P\delta_{ik})}{\partial x^k} = 0 ; \tag{6}$$

$$\frac{\partial (\rho E)}{\partial t} + \frac{\partial (\rho U^k h_0)}{\partial x^k} = 0 . \tag{7}$$

Here $U^1 = U, U^2 = V$ are the velocity components, $h_0 = (U^2 + V^2)/2 + h$, $h = \frac{\gamma}{\gamma - 1}\frac{P}{\rho} = \gamma e$, $e = \frac{RT}{\gamma - 1}$, $E = \left(e + \frac{1}{2}(U^2 + V^2)\right)$ are enthalpies and energies (per unit volume), $P = \rho RT$ is the state equation and $\gamma = C_p/C_v = 1.4$ is the specific heat ratio.

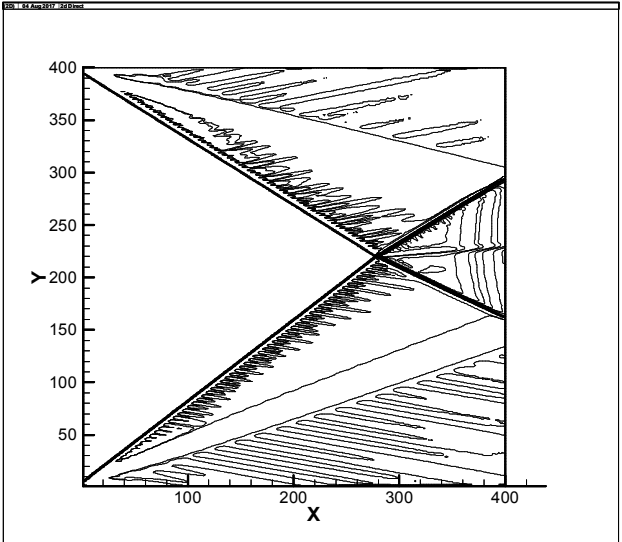 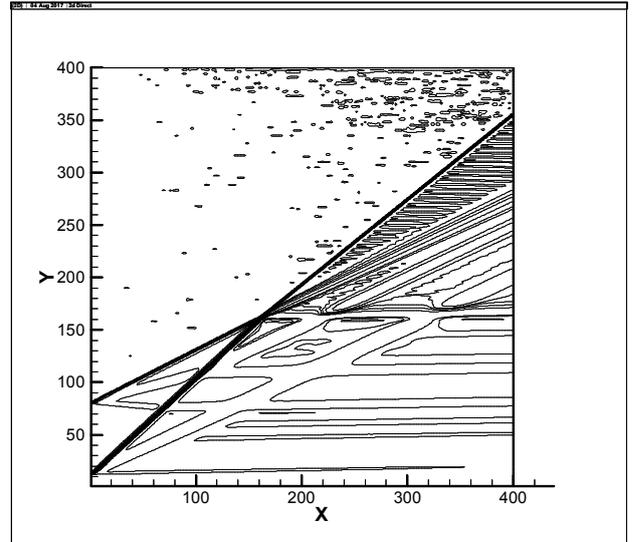

Fig. 1. Edney I density isolines.　　　　　Fig. 2. Edney VI density isolines.

The flow patterns, engendered by the single oblique shock wave and the interaction of shock waves of I and VI kinds according to Edney classification [39, 40], were used as the test problems due to the availability of analytic solutions. The flows were determined by the inflow and lateral boundaries conditions. The computations were performed for a Mach number range of $M = 3-5$, flow deflection angles range $\alpha = 10-30^o$. All tests correspond to the steady state solutions.

The values of analytical solution at grid points are considered as the exact solution. The flow field contains domains of continuous flow (nominal order of error is expected), shock waves (error order about $n = 1/2 \sim 1$ [2,3,5]), contact discontinuity line (error order about $n = 1/2$, [4]). As a result, one may hope to obtain a nontrivial error, composed of components with different orders of accuracy. The number of flow patterns with variable error order and available analytic solution is limited. All such two-dimensional steady flows are considered herein in order to estimate the approximation error magnitude.

Fig. 1 presents the computed isolines of density for Edney-I flow structure ($M = 3$ and flow deflection angles $\alpha_1 = 20^o$ and $\alpha_2 = 15^o$). The isolines are provided with a small step to illustrate the presence of a numerical error. The crossing shock waves and contact discontinuity line, engendered at the shocks crossing point, constitute the main elements of this flow structure.

Fig. 2 presents the density distribution for Edney-VI flow structure ($M = 4$, two consequent flow deflection angles $\alpha_1 = 10^o$, $\alpha_2 = 15^o$). The flow is determined by the merging shock waves, the contact line and the expansion fan.

### *5. Metrics used for comparison of flow fields*

Both *Theorem 1 and Conjecture 1* are stated for the distances determined by some metrics. The selection of the best metric is not evident. $L_1$ norm based metric seems to be most natural for problems related with shocks, since most results on approximation error are obtained in $L_1$ norm. However, most experience is related to the valuable functionals (lift, drag, etc). For norms engendered by some inner product ($L_2$, for example) the uncertainty of valuable functionals may be related to the error norm via the Cauchy–Bunyakovsky–Schwarz inequality

$$|\Delta \varepsilon| = |(\partial \varepsilon / \partial u \cdot (u_h - \tilde{u}))| < \|\partial \varepsilon / \partial u\|_{L_2} \cdot \|u_h - \tilde{u}\|_{L_2} \leq \|\partial \varepsilon / \partial u\|_{L_2} \cdot \|u^{(1)} - u^{(k)}\|_{L_2}. \qquad (8)$$

From this viewpoint, such norms may be more interesting in comparison with $L_1$.

Herein, we compare the metrics engendered by the $L_1, L_2, H^{-1}$ norms, $L_2$ based metrics which imitates a relative error (REM-$L_2$), and IMED metrics [37]. The metrics, having some physical meaning, illustrative capabilities, and a potential for flow comparison are not limited by above considerations, so the search for optimal metric is of further interest.

We consider the four component solution $u^{(i)} = \{\rho^{(i)}, U^{(i)}, V^{(i)}, e^{(i)}\}$. For the metrics engendered by the $L_1$ and $L_2$ norms, the distance between solutions is expressed as

$$\|u^{(i)} - u^{(k)}\|_{L_1} = \left(\frac{1}{N} \sum_{j=1}^{N} \left(\left|\rho^{(i)} - \rho^{(k)}\right|_j + \left|U^{(i)} - U^{(k)}\right|_j + \left|V^{(i)} - V^{(k)}\right|_j + \left|e^{(i)} - e^{(k)}\right|_j\right)\right), \qquad (9)$$

$$\|u^{(i)} - u^{(k)}\|_{L_2} = \left(\frac{1}{N} \sum_{j=1}^{N} \left((\rho^{(i)} - \rho^{(k)})_j^2 + (U^{(i)} - U^{(k)})_j^2 + (V^{(i)} - V^{(k)})_j^2 + (e^{(i)} - e^{(k)})_j^2\right)\right)^{1/2}. \qquad (10)$$

For CFD problems, the vector of solution contains elements having different physical meanings, such as density, velocity components, and energy. So, in parallel to Expressions (9,10), the distance between solutions was calculated using the normalized expression

$$\left\| \{ (\rho^{(i)} - \rho^{(k)}) / \left\| \rho^{(i)} \right\|, (U^{(i)} - U^{(k)}) / \left\| U^{(i)} \right\|, (V^{(i)} - V^{(k)}) / \left\| V^{(i)} \right\|, (e^{(i)} - e^{(k)}) / \left\| e^{(i)} \right\| \} \right\|_{L_2}, \qquad (11)$$

which imitates a relative error (we note this expression as "relative error metric" (REM-$L_2$)). It should be noted that expression (11) corresponds to the distance

$$(\Delta u^{(i)}, M \Delta u^{(i)}) = (M_{j,k} \Delta u_j^{(i)} \Delta u_k^{(i)})^{1/2}. \qquad (12)$$

This distance is determined by a diagonal metric tensor $M_{j,k}$ that describes some ellipsoid. With account of the presentation $M = A^* A$ (valid for a metric tensor as the symmetric positively defined matrix, a Mahalanobis distance metric [41]) one may state $(\Delta u^{(i)}, M \Delta u^{(i)}) = (\Delta u^{(i)}, A^* A \Delta u^{(i)})^{1/2} = (A \Delta u^{(i)}, A \Delta u^{(i)})^{1/2} = (\Delta z^{(i)}, \Delta z^{(i)})^{1/2}$. So, we can use the transformed space $Au^{(i)}$ (and corresponding $L_2$ norm) where the error may be described by a hypersphere.

The metric engendered by the Sobolev norm of negative order ($H^{-1}$) [42,43] also is of great interest due to the low regularity of the Euler equation solutions. According to [42], the Sobolev norm in $H^{-1}$ may be expressed as

$$\| f \|_{H^{-1}} = \sup_{\| u \|_{H^1} = 1} |(f, u)|. \qquad (13)$$

It was computed using the expression [43]
$$\| f \|_{H^{-1}} = (f, \tilde{u})_{L_2}, \qquad (14)$$

where $\tilde{u}$ is the solution of the screened Poisson equation

$$\lambda \frac{\partial^2 \tilde{u}}{\partial x^2} + \lambda \frac{\partial^2 \tilde{u}}{\partial y^2} - \tilde{u} = f. \qquad (15)$$

The coefficient $\lambda$ determines the smoothing properties for the transformation $\tilde{u} \to f$. The value of $\lambda$ was varied in the range $10^{-4} - 10^{-6}$. The calculations are performed by components for $f = \{ (\rho^{(i)} - \rho^{(k)}), (U^{(i)} - U^{(k)}), (V^{(i)} - V^{(k)}), (e^{(i)} - e^{(k)}) \}$ and corresponding $\tilde{u}$. We used the divergent integro-interpolation method [44] and the time relaxation approach to solve this equation.

The above considered metrics are sensitive to small variations of the flow field, such as shift of the shock wave location by single cell. Two numerical flow fields, engendered by such shift, are considered as distant and describing different flow structures. Thus, these distances do not capture the structural proximity between solutions. The Euclidean Distance, modified for analysis of images (IMage Euclidean Distance (IMED)), is of interest from this standpoint [37] since it provides some tolerance to shifts of solutions. It is described by the metric matrix

$$M_{ij} = \frac{1}{2\pi\sigma^2} \exp\{ -|P_i - P_j|^2 / (2\sigma^2) \}. \qquad (16)$$

The value $|P_i - P_j|$ is the distance between nodes $P_i$ and $P_j$ on the grid. For example, if $P_i$ corresponds to the cell $(k,l)$, and $P_j$ corresponds to the cell $(k_1, l_1)$, $|P_i - P_j|$ may be estimated as

$$|P_i - P_j| = ((k - k_1)^2 + (l - l_1)^2)^{1/2}. \qquad (17)$$

For two dimentional problem, the distance was estimated using the following form (presented here only for density)

$$(\Delta\rho^{(i)}, M\Delta\rho^{(i)}) = (\sum_{j,k,m,n} \frac{1}{2\pi\sigma^2} \exp\{-((j-m)^2 + (k-n)^2)/(2\sigma^2)\} \Delta\rho^{(i)}_{j,k} \Delta\rho^{(i)}_{m,n})^{1/2}. \qquad (18)$$

At $\sigma \leq 0.25$ the probability distribution approximation is of poor quality. At $\sigma = 0.5 - 1$ the values obtained by (18) are close to $L_2$ norm.

Asymptotically $H^{-1}$ and IMED tends to $L_2$ as $\lambda \to 0$ or $\sigma \to 0$.

### 6. Results of numerical tests

The analysis concerns the ensemble of computations performed by the following methods.

The first order scheme by [45] marked as $S1$ was used in the variant described by [46].

The second order scheme based on the MUSCL method [47] and using algorithm by [48] at cell boundaries is denoted as $S2$.

Second order TVD scheme of relaxation type by [49] is denoted as $S2TVD$.

Third order modified Chakravarthy-Osher scheme [50, 51] is marked as $S3$.

Fourth order scheme by [52] is marked as $S4$.

The FORTRAN codes by [49] are used for $S2TVD$. All other solvers were coded by the authors.

Computations were performed on uniform grids containing $100 \times 100$, $200 \times 200$ and $400 \times 400$ cells.

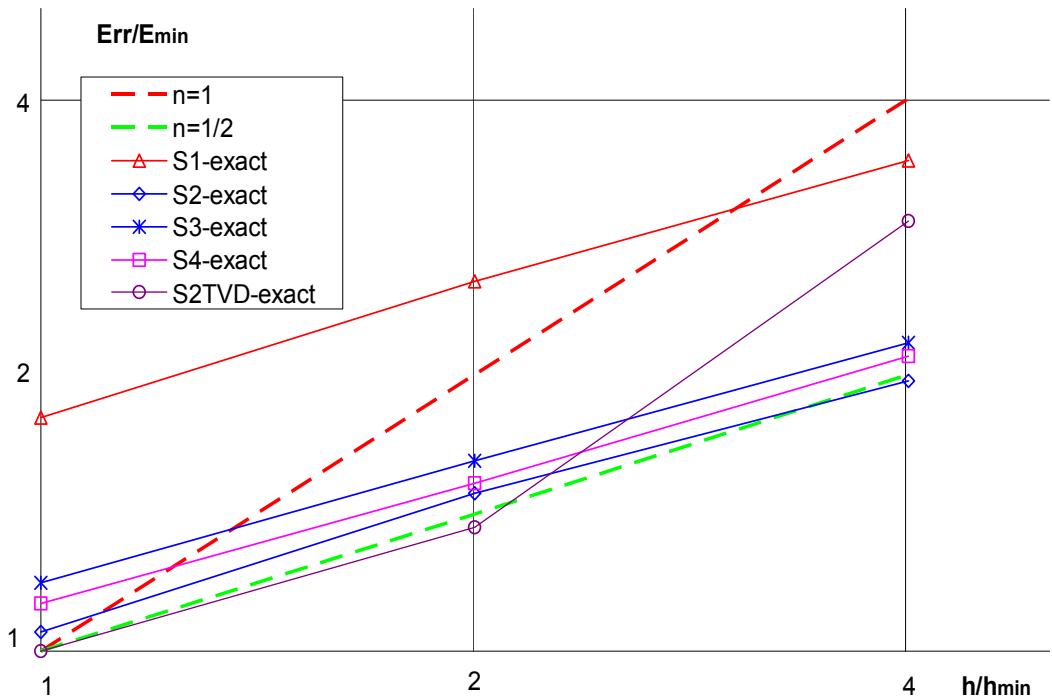

Fig. 3. Error norms in dependence on the step (logarithm scale)

Methods $S1, S2, S3, S4$ (1,2,3 and 4 nominal truncation orders) demonstrated the order of convergence slightly below $n = 1/2$ in norm $L_2$. In norm $L_1$ the same computations demonstrated the order of convergence slightly higher than $n = 1/2$. Second order $S2TVD$ scheme [49] from standpoint of error norm is close to first order scheme $S1$ for $100 \times 100$ grid and to high order schemes for grid $400 \times 400$. For illustration, $L_2$ error norms (normalized by the minimum value at finest grid) are given in Fig. 3 for considered grid levels plotted on a log-log scale (Edney-VI flow structure). Some pseudo asymptotic range is visible (the observed order of convergence is far from the nominal orders).

The calculations on the grid $100 \times 100$ demonstrated the formation of clusters in distances between $S2TVD$ and $S2, S3, S4$ solutions and successful estimation of the approximation error. However, the distances between solutions engendered by $S2TVD$ and $S2, S3, S4$ do not form clusters on the grid of $400 \times 400$. Paradoxically, the reason for this failure is the relatively rapid convergence of $S2TVD$. So, the results, obtained by comparison of $S1$ and $S2, S3, S4$ methods, are provided for illustration.

Since the relation of errors for chosen methods ($S1, S2, S3, S4$) is constant (due to close orders of convergence), the success of the triangle based error estimation does not depend on the step size. This situation is quite different from the case when convergence orders are nominal and error relation increases at grid refining that provides best chances for the triangle based method at fine grids.

We first check *Conjecture 1* and, second, verify the error estimation. It is considered as successful, if the error estimate $d(u^{(i)} - u^{(k)})$ is greater than the true error $d(u^{(k)} - \tilde{u})$, obtained in comparison with the analytical solution $\tilde{u}$.

The tests permit to conclude that the solutions obtained by the scheme $S1$ (as "inaccurate") and by $S2$, $S3$, $S4$ (as "accurate") enable us to find a vicinity of numerical solution that contains the exact solution for all tested grids.

The comparison of results obtained by schemes $S2, S3, S4$ does not enable to select clusters and to enclose the exact solution. These schemes produce solutions with errors which are close in magnitude and splitting into clusters is not observed.

If the *Conjecture 1* is not satisfied, the error estimation fails. However, the exact error value is about two or three maximum distances between numerical solutions, a result that provides an additional way for error estimation.

The numerical tests for the single oblique shock demonstrate the feasibility for error estimation, for example, see Tables 1 and 2. Table 1 demonstrates the formation of clusters in different metrics (S2-S4, S3-S4 and S3-S2 are smaller than S2-S1, S3-S1, S4-S1). Table 2 demonstrates the successful error estimation with exclusion of the small violation in $L_2$. The tests correspond to $M = 4$, flow deflection angle $\alpha_1 = 10^o$, and a grid of $100 \times 100$ mesh points.

Table 1. Distances between solutions for single shock test.

| $u^{(i)} - u^{(k)}$ | S4-S2 | S3-S2 | S3-S4 | S2-S1 | S3-S1 | S4-S1 |
|---|---|---|---|---|---|---|
| $L_1$ | 0.00569 | 0.0052 | 0.0032 | 0.0186 | 0.023 | 0.024 |
| $L_2$ | 0.0199 | 0.0287 | 0.011 | 0.0452 | 0.0566 | 0.060 |
| $H^{-1}$ | 0.0086 | 0.0076 | 0.00278 | 0.029 | 0.035 | 0.036 |
| REM-$L_2$ | 0.048 | 0.0145 | 0.022 | 0.0945 | 0.115 | 0.124 |
| IMED | 0.0145 | 0.0126 | 0.005 | 0.0519 | 0.067 | 0.0536 |

Table 2. Error estimation for single shock test.

| $u^{(i)} - u^{(k)}$ | S2-S1 | S2-exact | S3-S1 | S3-exact | S4-S1 | S4-exact |
|---|---|---|---|---|---|---|
| $L_1$ | 0.0186 | 0.0116 | 0.023 | 0.0092 | 0.024 | 0.0066 |
| $L_2$ | 0.0452 | 0.0459 | 0.0566 | 0.0407 | 0.060 | 0.0337 |
| $H^{-1}$ | 0.029 | 0.0137 | 0.035 | 0.00934 | 0.036 | 0.00784 |
| REM-$L_2$ | 0.0945 | 0.0923 | 0.115 | 0.079 | 0.124 | 0.0655 |
| IMED | 0.0519 | 0.0231 | 0.067 | 0.0155 | 0.0536 | 0.012 |

For Edney-VI shock interaction ($M = 4$, $\alpha_1 = 10^o$, $\alpha_2 = 15^o$, $100 \times 100$), the set of distances between solutions also splits into clusters. There is successful error estimation, see Tables 3-4.

Table 3. Distances between solutions for Edney-VI test.

| $u^{(i)} - u^{(k)}$ | S4-S2 | S3-S2 | S4-S3 | S2-S1 | S3-S1 | S4-S1 |
|---|---|---|---|---|---|---|
| $L_1$ | 0.023 | 0.0098 | 0.021 | 0.0668 | 0.072 | 0.0874 |
| $L_2$ | 0.059 | 0.025 | 0.051 | 0.149 | 0.16 | 0.191 |
| $H^{-1}$ | 0.028 | 0.0107 | 0.0127 | 0.0976 | 0.0928 | 0.121 |
| REM-$L_2$ | 0.051 | 0.0189 | 0.041 | 0.136 | 0.145 | 0.170 |
| IMED | 0.043 | 0.0195 | 0.035 | 0.179 | 0.192 | 0.171 |

The results of the generalized Richardson extrapolation [20,21] are presented in Table 4 for comparison sake. A set of solutions on consecutive higher resolution meshes ($100 \times 100$, $200 \times 200$, $400 \times 400$ cells) was used. The results of error norm estimation by RE are close to exact values. RE provides the error in the vector form that is the serious advantage, nevertheless, the required computer resources are much greater if compared with these required for the single grid approach.

Table 4. Error estimation for Edney-VI test.

| $u^{(i)} - u^{(k)}$ | S2-S1 | S2-exact | S3-S1 | S3-exact | S4-S1 | S4-exact |
|---|---|---|---|---|---|---|
| $L_1$ | 0.0668 | 0.046 | 0.072 | 0.046 | 0.0874 | 0.0375 |
| $L_2$ | 0.149 | 0.128 | 0.16 | 0.138 | 0.191 | 0.133 |
| $H^{-1}$ | 0.0976 | 0.055 | 0.0928 | 0.0603 | 0.121 | 0.0607 |
| REM-$L_2$ | 0.136 | 0.0898 | 0.145 | 0.093 | 0.170 | 0.0846 |
| IMED | 0.179 | 0.076 | 0.192 | 0.084 | 0.171 | 0.098 |
| $L_2$ (RE) | - | 0.139 | - | 0.137 | - | 0.141 |

Tables 5 and 6 present the results for different metrics from the viewpoint of error estimation for Edney-I test ($M = 3$, flow deflection angles $\alpha_1 = 20^o$ and $\alpha_2 = 15^o$, $100 \times 100$).

The tests by Tables 1-6 correspond to the relatively coarse mesh $100 \times 100$. A similar behavior is observed for finer grids ($200 \times 200$ and $400 \times 400$). Tables 7 and 8 present the results for different metrics from the viewpoint of error estimation for Edney-I test ($M = 3$, flow deflection angles $\alpha_1 = 20^o$ and $\alpha_2 = 15^o$, $400 \times 400$). Edney-I test is selected since it demonstrates the worst results when compared with the single shock and Edney-VI tests, respectively.

Table 5. Distances between solutions for Edney-I test. Coarse mesh.

| $u^{(i)} - u^{(k)}$ | S4-S2 | S3-S2 | S4-S3 | S2-S1 | S3-S1 | S4-S1 |
|---|---|---|---|---|---|---|
| $L_1$ | 0.017 | 0.018 | 0.019 | 0.0563 | 0.0673 | 0.0721 |
| $L_2$ | 0.044 | 0.043 | 0.045 | 0.107 | 0.128 | 0.141 |
| $H^{-1}$ | 0.0164 | 0.0154 | 0.0129 | 0.0609 | 0.0705 | 0.075 |
| REM-$L_2$ | 0.05 | 0.039 | 0.043 | 0.122 | 0.14 | 0.159 |
| IMED | 0.028 | 0.028 | 0.022 | 0.126 | 0.148 | 0.159 |

Table 6. Error estimation for Edney-I test. Coarse mesh.

| $u^{(i)} - u^{(k)}$ | S2-S1 | S2-exact | S3-S1 | S3-exact | S4-S1 | S4-exact |
|---|---|---|---|---|---|---|
| $L_1$ | 0.0563 | 0.0436 | 0.0673 | 0.050 | 0.0721 | 0.039 |
| $L_2$ | 0.107 | 0.124 | 0.128 | 0.146 | 0.141 | 0.139 |
| $H^{-1}$ | 0.0609 | 0.0512 | 0.0705 | 0.587 | 0.075 | 0.0597 |
| REM-$L_2$ | 0.122 | 0.163 | 0.14 | 0.178 | 0.159 | 0.176 |
| IMED | 0.126 | 0.114 | 0.148 | 0.0902 | 0.159 | 0.129 |

Table 7. Distances between solutions for Edney-I test. Fine mesh.

| $u^{(i)} - u^{(k)}$ | S4-S2 | S3-S2 | S4-S3 | S2-S1 | S3-S1 | S4-S1 |
|---|---|---|---|---|---|---|
| $L_1$ | 0.0061 | 0.0052 | 0.0068 | 0.0169 | 0.0202 | 0.0223 |
| $L_2$ | 0.0217 | 0.0226 | 0.0227 | 0.0545 | 0.0655 | 0.0709 |
| REM-$L_2$ | 0.026 | 0.020 | 0.022 | 0.0644 | 0.0739 | 0.0830 |
| $H^{-1}$ | 0.0148 | 0.0157 | 0.0135 | 0.0649 | 0.0764 | 0.0820 |
| IMED | 0.014 | 0.0145 | 0.011 | 0.0615 | 0.0764 | 0.0815 |

Table 8. Error estimation for Edney-I test. Fine mesh.

| $u^{(i)} - u^{(k)}$ | S2-S1 | S2-exact | S3-S1 | S3-exact | S4-S1 | S4-exact |
|---|---|---|---|---|---|---|
| $L_1$ | 0.0169 | 0.0122 | 0.0202 | 0.0147 | 0.0223 | 0.0123 |
| $L_2$ | 0.0545 | 0.0680 | 0.0655 | 0.0802 | 0.0709 | 0.0760 |
| REM-$L_2$ | 0.0644 | 0.0662 | 0.0739 | 0.0767 | 0.0830 | 0.0754 |
| $H^{-1}$ | *0.0458* | 0.0456 | 0.0546 | 0.0548 | 0.0577 | 0.0521 |
| IMED | 0.0615 | 0.0527 | 0.0764 | 0.0638 | 0.0815 | 0.0625 |

Tables 1-8 demonstrate that $L_1$ successfully performs the error estimation for all tests. $L_2$ and REM-$L_2$ fail for significant part of tests.

$H^{-1}$ engendered metric provides an intermediate quality. The IMED metric [37] enables the successful error estimation for most of the tests. So, the choice of metric proves to be crucial for the error magnitude estimation.

The tightness of estimation varies in dependence on the flow pattern. However, the violation of error estimates in these tests is moderate.

## *7. Discussion*

The relation of errors, obtained in above analysis, is not necessarily attributed to properties of considered schemes. It may be caused by the imperfections of numerical realization by the authors. Hence, the authors do not pretend to assess the considered numerical schemes. We are mainly concerned with the verification of the non-intrusive single-grid error estimator based on the numerical solutions obtained by the solvers of different accuracy.

The standard grid convergence analysis is based on the heuristic rule by C. Runge [13]. From this viewpoint, if the difference of two approximate solutions on the coarse grid and on the fine grid is small, then numerical solutions are close to exact solution. This rule is not applicable, if there is no grid convergence, the examples of such problems are provided by [10]. Also, this rule may be wrong if the convergence rate is slow. For example, [6,7] considers orders of convergence $n = 1/4 \div 1/6$ for multidimensional finite volume methods, while [8] considers elliptic boundary value problems, whose finite element approximations converge arbitrarily slow.

In a more rigorous approach, one should aim to obtain the error estimate of form $\delta(u_h, \tilde{u}) \leq \delta$ with computable $\delta$. Formally, the Richardson method [17-21] is close to this ideal. It enables to determine the refined solution and the error estimate, if the single error order exists in the total flow field. The set of two solutions, computed for different meshes, is used. Unfortunately, in most CFD problems the error order on different flow structures varies [2-8]. The estimation of the local order and the check of the asymptotic range need four consequent meshes (at least) that causes the significant requirements for computer resources [21].

We considered a single-grid supplement to the Richardson method and Runge rule, based on the ensemble of solutions obtained by different solvers. The above considered method may be used away from the asymptotic range as a postprocessor. A mesh refinement should be performed only if the magnitude of the error is not acceptable.

The feasibility of estimating the distance from the exact solution to numerical one seems to be attractive. However, the numerical value of a threshold, when two approximate solutions can be considered as describing the same flow, may not be evident. Also, the magnitude of the global error ($\delta(u_h, \tilde{u})$ or norm $\|u_h - \tilde{u}\|$) is not very informative in CFD, since most experience is related with the valuable goal functionals (lift, drag, etc). From this viewpoint, the global error may be related to the uncertainty of some valuable functionals via the Cauchy–Bunyakovsky–Schwarz inequality, if the error magnitude estimate is engendered by some inner product (that restricts the range of $L_1$ norm applicability).

The existence of "accurate" and "inaccurate" schemes is one of the most important notions of the computational mathematics, unfortunately, it is usually defined in the asymptotic sense. The above numerical results demonstrate the feasibility of distinguishing between "accurate" and "inaccurate" solutions in the sense of error ranging in certain metric. For example, the distributions of distances between solutions provided in Tables 1,3,5,7 show the presence of two clusters corresponding to "accurate" and "inaccurate" solutions. This engenders the hope to estimate the upper error bound only from observable values of distances between solutions (without *a priori* information on errors ranging), a hope that is confirmed by Tables 2,4,6,8.

If there is no splitting into clusters, the maximum distance between solutions provides the possibility for a rough estimation of numerical error, since it is relatively close to the distance between numerical and analytical solutions.

We consider herein only regular uniform grids omitting the consideration the irregular grids, for which the truncation error order is less the nominal value [9].

At first glance, the present approach is similar to the "p-refinement", widely used in the domain of finite elements [53]. However, "p-refinement" cannot be applied in situations (typical for flows with discontinuities) when the order of error does not depend on the choice of algorithm. There exists some version of Richardson extrapolation [54], which utilizes three finite element solutions with consecutive orders of accuracy that has, at first glance, some analogy with our technique. However, algorithm [54] is based on specific asymptotic of energy norms and is not related with the triangle inequality and formation of clusters of solutions.

The dependence on a choice of numerical methods, the analyzed solution, and the metric is the drawback of considered ensemble based method. The same set of methods may provide the segregation into clusters for one flow pattern (or grid size) and may not provide it for another. So, this approach cannot replace the standard verification (by mesh refining or by Richardson extrapolation) and is aimed to supplement it by a fast non-intrusive algorithm.

## *8. Conclusions*

The approximation error estimation is feasible, if the set of distances between numerical solutions, obtained by independent solvers, is split into separated clusters corresponding to "accurate" and "inaccurate" solutions. The distance between clusters should exceed the dimension of "accurate" cluster.

The numerical tests confirmed the applicability of this heuristic rule for two dimensional supersonic steady problems, governed by Euler equations, with the dependence on the choice of the metric.

The $L_1$ based metric operates successfully in all tests.

The $L_2$ based metric and a metric which imitates the relative error (REM-$L_2$), fail rather often, albeit with moderate violations.

The metric, engendered by $H^{-1}$ norm, provides an intermediate reliability.

The IMED [37] metric demonstrated the quality of the error estimation comparable with the $L_1$ based metric.

## *Acknowledgement*

Authors acknowledge the partial support by grants of RFBR № 16-01-00553A and 17-01-444A.